 \documentclass[preprint]{aastex}
\shorttitle{Magnetic Accretion in GX\,301--2}
\shortauthors{Ikhsanov \& Finger}
\usepackage{graphicx}
\usepackage[mathscr]{eucal}

\newcommand{\be}{\begin{equation}}
\newcommand{\ee}{\end{equation}}
\newcommand{\bdm}{\begin{displaymath}}
\newcommand{\edm}{\end{displaymath}}
\def\dmf{\dot{\mathfrak{M}}}
\def\msE{\mathscr{E}}
\newcommand{\vect}[1]{\mathbf{#1}}
\begin{document}

\title{Signs of Magnetic Accretion in the X-ray Pulsar Binary GX\,301--2}

\author{Nazar\,R.\,Ikhsanov}
\affil{Central Astronomical Observatory of the Russian Academy of Sciences at Pulkovo, 196140 St.\,Petersburg, Russia}

\author{Mark\,H.\,Finger\altaffilmark{1}}
\affil{Universities Space Research Association, 6767 Old Madison Pike, Suite 450, Huntsville, AL 35806, USA}
\altaffiltext{1}{National Space Science and Technology Center, 320 Sparkman Drive, Huntsville, AL}

\begin{abstract}
Observations of the cyclotron resonance scattering feature in the X-ray spectrum of GX\,301--2 suggest that the surface field of the neutron star is $B_{\rm CRSF} \sim 4 \times 10^{12}$\,G. The same value has been derived in modelling the rapid spin-up episodes in terms of the Keplerian disk accretion scenario. However, the spin-down rate observed during the spin-down trends significantly exceeds the value expected in currently used spin-evolution scenarios. This indicates that either the surface field of the star exceeds $50\,B_{\rm CRSF}$, or a currently used accretion scenario is incomplete. We show that the above discrepancy can be avoided if the accreting material is magnetized. The magnetic pressure in the accretion flow increases more rapidly than its ram pressure and, under certain conditions, significantly affects the accretion picture. The spin-down torque applied to the neutron star in this case is larger than that evaluated within a non-magnetized accretion scenario. We find that the observed spin evolution of the pulsar can be explained in terms of the magnetically controlled accretion flow scenario provided the surface field of the neutron star is $\sim B_{\rm CRSF}$.
\end{abstract}


\section{Introduction}

A strong magnetization of the neutron star in the High Mass X-ray Binary (HMXB) GX\,301--2 has first been suspected by \citet{Lipunov-1982}. Assuming the star to rotate at the equilibrium period given by \citet{Davidson-Ostriker-1973} he limited its dipole magnetic moment to $\ga 10^{32}\,{\rm G\,cm^3}$.This implies the surface field of the star to be in excess of $2 \times 10^{14}$\,G.

This estimate has been challenged by observations of the cyclotron resonance scattering feature (CRSF) at $\sim 35-45$\,keV \citep[see][and references therein]{La-Barbera-etal-2005}. The field strength of the pulsar inferred from these observations, $B_{\rm CRSF} \sim 4 \times 10^{12}$\,G, is a factor of 50 smaller than the surface field evaluated by \citet{Lipunov-1982}.

However, a question about the field strength of GX\,301--2 has recently been raised again by  \citet{Doroshenko-etal-2010} who reported the pulsar to spin down on a time scale of years at the average rate of $\dot{P}_{\rm s} \simeq 4.25 \times 10^{-8}\,{\rm s\,s^{-1}}$ ($\dot{\nu}_{\rm sd} = \dot{P}_0/P_{\rm s}^2 \sim -10^{-13}\,{\rm Hz\,s^{-1}}$). Using the spin-down scenarios of \citet{Davidson-Ostriker-1973} and \citet{Illarionov-Kompaneets-1990} they put the same lower bound on the surface field of the neutron star as \citet{Lipunov-1982}. To avoid the apparent discrepancy they have suggested that CRSF is generated at the top of a radiative-dominated accretion column, which is extended above the stellar surface up to an altitude of a few stellar radii.

\begin{figure}
\centerline{\includegraphics[width=9cm]{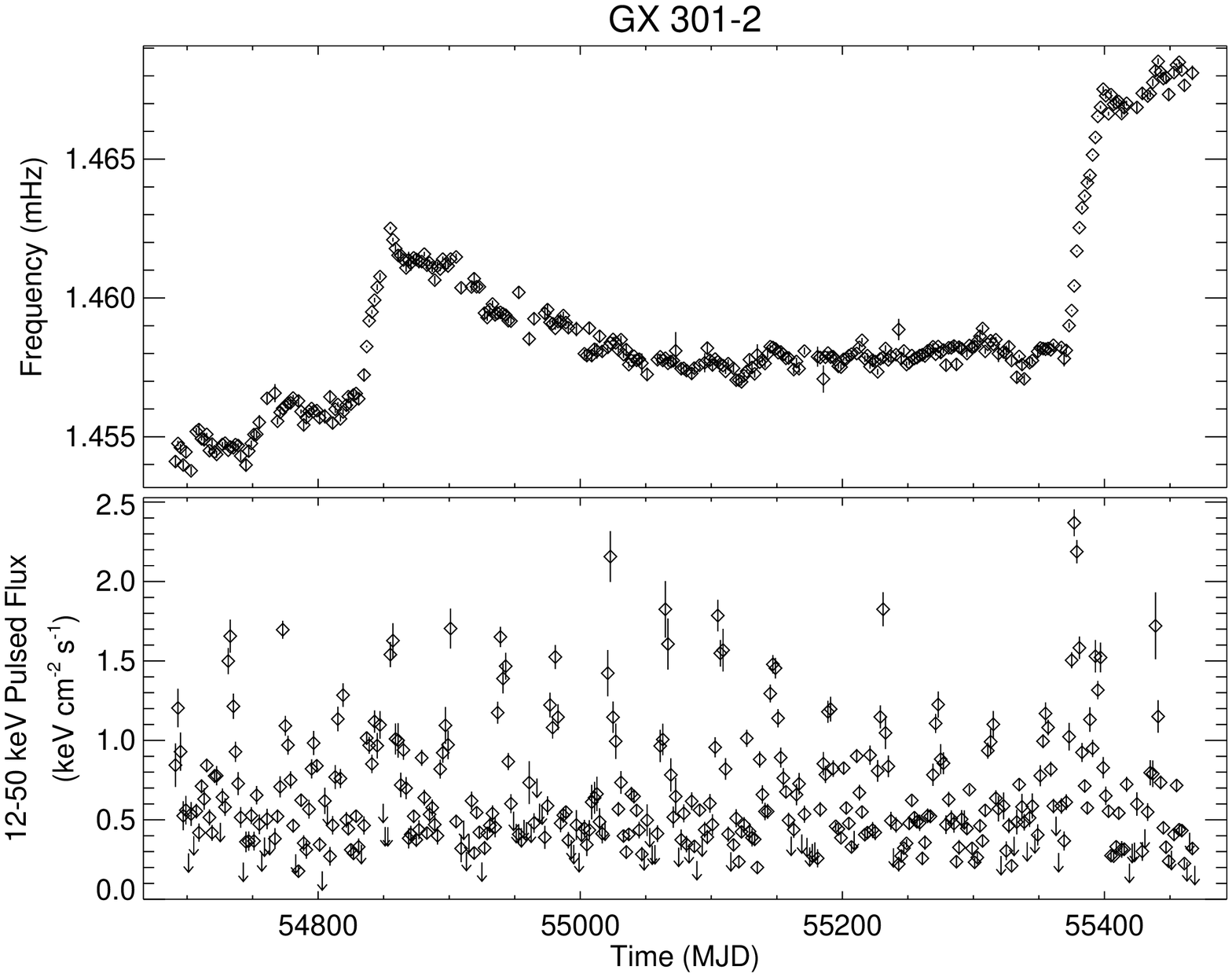}}
\caption{Pulse frequency and 12--50\,keV pulsed flux observed with the Fermi Gamma-ray Burst
Monitor detectors. GMB Pulsar Project http://gammaray.nsstc.nasa.gov/gbm/science/pulsars/
\label{light-curve}}
\end{figure}

In this paper we argue that this conclusion is premature. The spin behavior of the pulsar is very complicated. The long-term spin-down trends are superposed with phases of stable rotation and rapid spin-up episodes (see Figure\,\ref{light-curve}, and Sect.\,\ref{basic}). The value of torques applied to the star strongly depends on the geometry and physical conditions in the accreting material (Sect.\,\ref{adse}). The pulsar behavior during the rapid spin-up episodes can be explained in terms of accretion from a Keplerian disk \citep{Koh-etal-1997} provided the surface field of the star is $\sim B_{\rm CRSF}$ (Sect.\,\ref{spin-up}). The spin-down rate of the star during the spin-down trends strongly depends on the conditions in the material surrounding the pulsar magnetosphere (Sect.\,\ref{spin-down}). We find that the neutron star can brake harder if the accreting material is magnetized. The accretion flow in this case is approaching the magnetosphere of the neutron star in a form of a dense magnetic slab confined by the magnetic field of the flow itself. The interaction between the slab and stellar magnetosphere leads to an effective angular momentum transfer from the star to the slab (Sect.\,\ref{mca}). We show that the observed spin-down rate can be explained within this magnetically controlled accretion scenario provided the surface field of the neutron star is about $B_{\rm CRSF}$ (Sect.\,\ref{magx}). Assumptions adopted in this accretion picture are briefly discussed in Sect.\,\ref{discussion} and our conclusions are summarized in Sect.\,\ref{concl}.

 \section{Basic parameters}\label{basic}

GX\,301--2 is a 685\,s accreting X-ray pulsar in a 41.5\,d eccentric orbit around the early B-type supergiant companion Wray\,977 \citep{Sato-etal-1986}. The massive companion underfills its Roche lobe and looses material at the rate $\dot{M}_{\rm out} \simeq 10^{-5}\,{\rm M_{\sun}\,yr^{-1}}$ in the form of a relatively slow ($v_{\rm w} \sim 300-400\,{\rm km\,s^{-1}}$) stellar wind \citep{Kaper-etal-2006}.

As the neutron star moves through the wind of density $\rho_{\infty}$ with a relative velocity $\vect{v}_{\rm rel} = \vect{v}_{\rm ns} + \vect{v}_{\rm w}$ it captures material at a rate $\dmf \leq \dmf_{\rm c}$, where
 \be
\dmf_{\rm c} = \pi R_{\rm G}^2 \rho_{\infty} v_{\rm rel}.
 \ee
Here $R_{\rm ns}$ is the radius, $M_{\rm ns}$ the mass and $v_{\rm ns} \sim 250\,{\rm km\,s^{-1}}$ is the orbital velocity of the neutron star, and $R_{\rm G} = 2GM_{\rm ns}/v_{\rm rel}^2$ is its Bondi radius.

The interaction between the accreting material and the stellar magnetic field leads to formation of a magnetosphere which, in the first approximation, prevents the accretion flow from reaching the stellar surface. The flow is decelerated by the stellar magnetic field at a distance $r_{\rm m}$, which is defined by equating the pressure of the accretion flow with the magnetic pressure due to the dipole field of the neutron star and is referred to as the radius of the magnetosphere. The accretion flow enters the pulsar field at the magnetospheric boundary and flowing along the field lines reaches the surface of the neutron star at the magnetic pole regions. The rate at which the accreting material enters the pulsar field, $\dmf_{\rm in}$, and, correspondingly, the mass accretion rate onto the stellar surface, $\dmf_{\rm a}$, can be evaluated as
 \be
\dmf_{\rm in} = \dmf_{\rm a} \simeq 10^{17} R_6 m^{-1} \left(\frac{L_{\rm X}}{2 \times 10^{37}\,{\rm erg\,s^{-1}}}\right)\,{\rm g\,s^{-1}},
 \ee
where $R_6 = R_{\rm ns}/ 10^6$\,cm, $m=M_{\rm ns}/1.4\,{\rm M_{\sun}}$ and $L_{\rm X}$ is the X-ray luminosity of the system, which is normalized to its average value $L_{\rm X}\sim (1-3) \times 10^{37}\,{\rm erg\,s^{-1}}$ for a distance of 1.8--3\,kpc, \citep{Chichkov-etal-1995, Kaper-etal-2006}.

The persistent character of the pulsar indicates that the magnetospheric radius of the neutron star is smaller than its corotation radius,
 \be\label{rcor}
r_{\rm cor} = \left(\frac{GM_{\rm ns}}{\omega_{\rm s}^2}\right)^{1/3} \simeq 1.3 \times 10^{10}\ m^{1/3}\ P_{685}^{2/3}\ {\rm cm},
 \ee
and hence, the centrifugal barrier at the magnetospheric boundary does not prevent the accretion flow from reaching the stellar surface \citep{Shvartsman-1970}. Here $P_{685}$ is the spin period of the neutron star in units of 685\,s, and $\omega_{\rm s} = 2 \pi/P_{\rm s}$ is its angular velocity.

The pulsar shows a complicated picture of spin variation (see Figure\,\ref{light-curve}), which can be roughly divided into three phases. The spin-down trends with the average rate $|\dot{\nu}_{\rm d0}| \sim 10^{-13}\,{\rm Hz\,s^{-1}}$ and 1--3~years duration \citep{Doroshenko-etal-2010} are superposed with the short-term (about 30~days) rapid spin-up episodes with $\dot{\nu}_{\rm u0} \simeq 5 \times 10^{-12}\,{\rm Hz\,s^{-1}}$ \citep{Koh-etal-1997}, and the phase of almost stable rotation ($\dot{\nu} < 10^{-13}\,{\rm Hz\,s^{-1}}$), which lasts up to a few years \citep{Bildsten-etal-1997}. The pulsar transition between these phases occurs without any significant changes in the intensity and spectral parameters of the X-ray emission. In particular, X-ray luminosity of the system during rapid spin-up episodes is only a factor of 2--4 larger than that observed during the spin-down trends \citep{Koh-etal-1997}. This indicates that the mass-transfer between the system components operates almost stationary (i.e. $\dmf_{\rm c} \sim \dmf_{\rm a} = \dmf$) without any significant changes in the parameters of the accretion flow.

  \section{Accretion-driven spin evolution}\label{adse}

The equation governing the spin evolution of an accreting neutron star reads
  \be\label{main}
I \dot{\omega}_{\rm s} = K_{\rm su} - K_{\rm sd}.
 \ee
Here $I$ is the moment of inertia of the neutron star, and $K_{\rm su}$ and $K_{\rm sd}$ are the spin-up and spin-down torques applied to the star from the accretion flow.
 \subsection{Spin-up torque}

Since the material captured by the neutron star in a binary system possesses specific angular momentum, the mass accretion onto the neutron star is accompanied by the accretion of angular momentum at the rate $\dot{J} = \xi\,\Omega_{\rm orb}\,R_{\rm G}^2\, \dmf_{\rm c}$ \citep[for discussion see, ][and references therein]{Illarionov-Sunyaev-1975, Illarionov-Kompaneets-1990}. Here $\Omega_{\rm orb} = 2 \pi/P_{\rm orb}$ is the orbital angular velocity and $\xi$ is a parameter accounting for dissipation of angular momentum in the accreting material \citep[see e.g.][and references therein]{Ruffert-1999}.

The geometry of the accretion flow in this case deviates from spherical symmetry. The deviation, however, will be significant only if $r_{\rm circ} \ga r_{\rm m}$, where $r_{\rm circ} = \dot{J}^2/GM_{\rm ns} \dmf_{\rm c}^2$ is a circularization radius, at which the angular velocity of the accreting material, $\omega_{\rm en} = \xi\,\Omega_{\rm orb}\,\left(R_{\rm G}/r \right)^2$, reaches the Keplerian angular velocity, $\omega_{\rm k} = \left(r^3/2GM_{\rm ns}\right)^{1/2}$ \citep{Bisnovatyi-Kogan-1991}. If this condition is satisfied a formation of the Keplerian disk can be expected. Otherwise, the accretion should be treated in the Quasi-Spherical (QS) approximation, which implies $0 < \omega_{\rm en}(r_{\rm m}) \ll \omega_{\rm k}(r_{\rm m})$.

The radius of the magnetosphere of a neutron star accreting from a non-magnetized wind (validity of this approximation is discussed in Sect.\,\ref{mca}) is $r_{\rm m} \sim r_{\rm a}$, where \citep[see, e.g.][and references therein]{Arons-1993}
  \be\label{ra}
 r_{\rm a} = \left(\frac{\mu^2}{\dmf (2 GM_{\rm ns})^{1/2}}\right)^{2/7}.
 \ee
Here $\mu$ is the dipole magnetic moment of the neutron star. Solving inequality $r_{\rm circ} \ga r_{\rm a}$ for $v_{\rm rel}$ one finds that the neutron star in a wind-fed HMXB would be surrounded by a persistent Keplerian disk if $v_{\rm rel} \la v_{\rm cr}$, where \citep[see e.g.,][]{Ikhsanov-2007},
  \be\label{vcr}
v_{\rm cr} \simeq 200\ \xi_{0.2}^{1/4} \mu_{30}^{-1/14} m^{11/28} \dmf_{17}^{1/28} \left(\frac{P_{\rm orb}}{41.5\,{\rm d}}\right)^{-1/4} {\rm km\,s^{-1}}.
 \ee
Here $\mu_{30}=\mu/10^{30}\,{\rm G\,cm^3}$, $\dmf_{17} = \dmf/10^{17}\,{\rm g\,s^{-1}}$ and $\xi_{0.2} = \xi/0.2$ is normalized to its average value derived in numerical modeling of quasi-spherical accretion in the non-magnetized flow approximation \citep[][and references therein]{Ruffert-1999}. The specific angular momentum of the material in the Keplerian disk is $j_{\rm k}(r) = \omega_{\rm k} r^2$. Therefore, the spin-up torque associated with the accretion of material from the inner radius of the disk is $K_{\rm su}^{\rm (d)} = \dmf j_{\rm k}(r_{\rm m}) r_{\rm m}^2 = \left(GM_{\rm ns} r_{\rm m}\right)^{1/2}$ \citep{Pringle-Rees-1972}.

Thus, the spin-up torque applied to an accreting neutron star in a wind-fed HMXB can be expressed as
 \be\label{ksugen}
 K_{\rm su} \simeq \left\{
 \begin{array}{llll}
 \dmf\ \left(GM_{\rm ns} r_{\rm m}\right)^{1/2} & \mbox{for} & v_{\rm rel} \leq v_{\rm cr} & \mbox{(disk)} \\
 & & & \\
 \dmf\ \xi\,\Omega_{\rm orb}\,R_{\rm G}^2 & \mbox{for} & v_{\rm rel} > v_{\rm cr} & \mbox{(QS)}\\
  \end{array}
 \right.
 \ee

 \subsection{Spin-down torque}

The spin-down torque applied to an accreting neutron star is associated with interaction between its magnetic field and material surrounding the magnetosphere. This interaction leads to a distortion of the magnetospheric field, turbulization of the material at the magnetospheric boundary, excitation of MHD waves and magnetic reconnection. The efficiency of these processes depends on the field geometry, conductivity of the accreting material and spectrum of the turbulent motions. A detailed modeling of these processes is, therefore, very complicated and is beyond the scope of this paper. Here we focus on a simplified task targeting at evaluation of upper limits to the spin-down torque under various assumptions about the geometry and parameters of the accretion flow.

The spin-down torque applied to a neutron star accreting material from a Keplerian disk has first been evaluated by \citet{Lynden-Bell-Pringle-1974}. The stellar magnetosphere in this case contains both closed and open field lines, which are extended beyond the inner radius of the disk. The angular velocity of the material in those parts of the Keplerian disk, which are located inside the corotation radius, exceeds the angular velocity of the star itself. The interaction between these parts of the disk and stellar magnetosphere tends, therefore, to spin-up the star. The angular velocity of the material in the Keplerian disk is, however, decreasing with radius and becomes smaller than the angular velocity of the neutron star in the region $r > r_{\rm cor}$. The interaction between the magnetospheric field lines and the disk at this distance tends to spin-down the star. As shown by \citet{Lynden-Bell-Pringle-1974}, the spin-down torque associated with this interaction can be evaluated as $K_{\rm sd}^{\rm (d)} = k_{\rm t} \mu^2/r_{\rm cor}^3$, where $k_{\rm t} < 1$ is the efficiency parameter.

The magnetosphere of a neutron star undergoing spherical accretion is closed. The stellar field is screened by the surface currents at the magnetospheric boundary and there is no field lines extended beyond the magnetospheric radius. The boundary is convex towards the accreting material and is closed to the cusp points located at the stellar magnetic axis \citep[for discussion see][]{Arons-Lea-1976, Elsner-Lamb-1977, Michel-1977}. This indicates that the interaction between the stellar magnetosphere and the accretion flow
occurs solely at the the magnetospheric boundary. For the neutron star in this case to spin-down the angular velocity of the accreting material at the magnetospheric boundary, $\omega_{\rm en}(r_{\rm m})$, should be
smaller than the angular velocity of the star itself, $\omega_{\rm s}$ \citep{Bisnovatyi-Kogan-1991}. We assume here that this condition is satisfied and discuss the validity of this assumption in Sect\,\ref{spin-down}.

The spin-down torque applied to a neutron star accreting form a quasi-spherical accretion flow, in the first approximation, can be evaluated by finding the torque on  a rotating sphere of the radius $r_{\rm m}$ in a viscous medium \citep[for discussion see e.g.,][]{Lipunov-1992}. The spin-down torque in this case can be expressed as
\be\label{ksd0}
 K_{\rm sd} = 4 \pi r_{\rm m}^2 \nu_{\rm t} \rho_0 v_{\phi},
 \ee
where $\nu_{\rm t}$ is the viscosity coefficient, $\rho_0$ is the density of material at the magnetospheric boundary and $v_{\phi} = \left[\omega_{\rm s} - \omega_{\rm en}(r_{\rm m})\right]\,r_{\rm m}$ is the azimuthal component of the relative linear velocity between the sphere and the surrounding material. For a simplicity we consider the case of $\omega_{\rm en}(r_{\rm m}) \ll \omega_{\rm s}$ and hence, $v_{\phi} = \omega_{\rm s}\,r_{\rm m}$. This approximation is reasonable for evaluation of upper limits to the spin-down torque.

Assuming a turbulent nature of the viscosity, $\nu_{\rm t} = k_{\rm t} v_{\rm t} \ell_{\rm t}$, and taking into account that the density of the free-falling material at the magnetospheric boundary of the neutron star is $\rho_0 = \dmf_{\rm c}/4 \pi r_{\rm m}^2 v_{\rm ff}(r_{\rm m})$, one finds
\be\label{ksd1}
 K_{\rm sd} = k_{\rm t}\,\omega_{\rm s}\,r_{\rm m}\,\ell_{\rm t}(r_{\rm m})\,\dmf_{\rm c}\,\frac{v_{\rm t}(r_{\rm m})}{v_{\rm ff}(r_{\rm m})}.
 \ee
Here $v_{\rm t}(r_{\rm m})$ and $\ell_{\rm t}(r_{\rm m})$ are the velocity and the scale of turbulent motions, $v_{\rm ff}(r_{\rm m}) = \left(2 GM_{\rm ns}/r_{\rm m}\right)^{1/2}$ is the free-fall velocity at the magnetospheric boundary and $k_{\rm t} \leq 1$ is the efficiency parameter. If the turbulent motions in the material at the magnetospheric boundary are excited by its interaction with the rotating magnetosphere, the viscosity coefficient can be expressed as $\nu_{\rm t}^{(0)} = k_{\rm t} v_{\phi} r_{\rm m}$. This implies that the scale and the velocity of turbulent motions are limited to $\ell_{\rm t} \leq r_{\rm m}$ and $v_{\rm t} \leq v_{\phi}$. Putting these parameters to Eq.~(\ref{ksd1}) and setting $r_{\rm m} = r_{\rm a}$ one finds
 \be
  K_{\rm sd}^{(0)} = k_{\rm t}\,\frac{\mu^2\,\omega_{\rm s}^2}{GM_{\rm ns}} = k_{\rm t}\,\frac{\mu^2}{r_{\rm cor}^3}.
 \ee
This upper limit to the spin-down torque has previously been evaluated by \citet{Lipunov-1982} and \citet{Bisnovatyi-Kogan-1991} for the case of quasi-spherical accretion.

The neutron star under the same conditions is braking harder if $\nu_{\rm t} > \nu_{\rm t}^{(0)}$. This can be realized if $v_{\rm t} > v_{\phi}$ (note, that the scale parameter, $\ell_{\rm t}$, has already been limited above to its maximum possible value). On the other hand, the velocity of turbulent motions is limited to the speed of sound in the accreting material since the supersonic turbulence it effectively suppresses by the Landau damping \citep[for discussion see, e.g.,][]{Davies-Pringle-1981}. Therefore, the maximum possible value of $\nu_{\rm t}$  can be found by setting $v_{\rm t} \sim v_{\rm ff}$, i.e. assuming that the velocity of the turbulent motions in the accretion flow is close to the speed of sound in the material heated in the shock at the magnetospheric boundary up to the adiabatic temperature \citep[for discussion see, e.g.][]{Arons-Lea-1976, Elsner-Lamb-1977}. Putting this to Eq.~(\ref{ksd1}) yields
 \be
K_{\rm sd}^{\rm (t)} = k_{\rm t}\,\dmf_{\rm c}\,\omega_{\rm s}\,r_{\rm m}^2.
 \ee
Finally, taking into account that $\dmf = \mu^2/\left(2\,r_{\rm m}^7 GM_{\rm ns}\right)^{1/2}$ (see Eq.~\ref{ra}), one can express the maximum possible spin-down torque as
 \be
 K_{\rm sd}^{\rm (t)} = k_{\rm t}\,\frac{\mu^2}{\left(r_{\rm m}\,r_{\rm cor}\right)^{3/2}}.
 \ee

The condition $v_{\rm t} > v_{\phi}$ can be satisfied if the neutron star accretes material from either a hot turbulent atmosphere \citep[as it occurs in the subsonic propeller scenario of][]{Davies-Pringle-1981}, or a magnetized flow with a high value of the magnetic viscosity. The subsonic propeller scenario can be used, however, for interpretation of a limited number of X-ray pulsars. The spin period of subsonic propellers is limited to $P_{\rm s} < P_{\rm br}$, where \citep{Ikhsanov-2001}
  \be\label{pbr}
P_{\rm br} \simeq\ 450\ \mu_{30}^{16/21}\ \dmf_{15}^{-5/7}\ m^{-4/21}\ {\rm s}
  \ee
is a so-called break period. If the spin period of a neutron star exceeds $P_{\rm br}$ the cooling of the  atmosphere due to bremsstrahlung emission and turbulent motions becomes more effective than heating (due to propeller action by the neutron star). The X-ray source in this case would be observed as an X-ray burster \citep{Lamb-etal-1977}. Furthermore, the X-ray luminosity of subsonic propellers is limited to $L_{\rm X} < L_{\rm br}$, where
 \be
 L_{\rm br} \simeq 10^{34}\ \mu_{30}^{8/7} m^{-2/7} \dmf_{15}^{-4/7} \left(\frac{P_{\rm s}}{500\,{\rm s}}\right)^{-1}\ {\rm erg\,s^{-1}}.
 \ee
Otherwise, cooling of the material at the magnetospheric boundary due to inverse Compton scattering of the X-ray photons, emitted from the surface of the neutron star, on the hot atmospheric electrons dominates heating due to propeller action. This condition has been derived by solving inequality $t_{\rm c}(r_{\rm m}) > t_{\rm h}$, where
 \be\label{tcomp}
t_{\rm c}(r) = \frac{3 \pi\,r^2\,m_{\rm e}\,c^2}{2\,\sigma_{\rm T}\,L_{\rm X}}
 \ee
is the Compton cooling time \citep{Elsner-Lamb-1977} and $t_{\rm h}  \sim 1/\omega_{\rm s}$ is the heating time due to propeller action \citep{Davies-Pringle-1981}. Here $m_{\rm e}$ is the electron mass and $\sigma_{\rm T}$ is the Thomson cross-section. Thus, the subsonic propellers can only appear as relatively low luminous, fast rotating pulsars, which is obviously not the case of GX\,301--2.

A high viscosity can also be expected if the material surrounding the neutron star is magnetized. The magnetic viscosity coefficient in this case can be expressed as $\nu_{\rm m} = k_{\rm t} V_{\rm A} \ell_{\rm m}$, where $V_{\rm A} = B_{\rm f}/(4 \pi \rho_0)^{1/2}$ is the Alfv\'en velocity and $\ell_{\rm m}$ is the characteristic scale of the magnetic field, $B_{\rm f}$, in the material at the magnetospheric boundary. A rapid braking of the neutron star in this case can be expected if $\ell_{\rm m} \sim r_{\rm m}$ and $V_{\rm A} > V_{\phi}$. The Magnetically Controlled Accretion scenario in which this situation is realized will be discussed in Sect.\,\ref{mca}.

Concluding this section one can express the spin-down torque applied to a neutron star from the accretion flow in the following form
\be\label{ksd00}
 K_{\rm sd} = \left\{
 \begin{array}{lll}
 k_{\rm t}\ \displaystyle\frac{\mathstrut \mu^2}{r_{\rm cor}^3} ~ & ~ \mbox{for} ~ & ~ \nu_{\rm t}\ \leq \ \nu_{\rm t}^{(0)}  \\
 & &  \\
k_{\rm t}\ \displaystyle\frac{\mathstrut \mu^2}{\left(r_{\rm m}\,r_{\rm cor}\right)^{3/2}} ~ & ~ \mbox{for} ~ & ~ \nu_{\rm t}\ >\ \nu_{\rm t}^{(0)} \\
  \end{array}
 \right.
 \ee
where $k_{\rm t} < 1$ is a parameter accounting for the conductivity, spectrum of turbulence and inhomogeneities of the accreting material.

    \section{Rapid spin-up episodes}\label{spin-up}

The spin-up torque applied to the neutron star during the rapid spin-up episodes is  $K_{\rm su} = K_0 + K_{\rm sd}$, where
 \be
 K_0 = 2 \pi I \dot{\nu}_{\rm su} \simeq 3 \times 10^{34}\ I_{45} (\dot{\nu}_{\rm su}/\dot{\nu}_{\rm u0})\,{\rm dyne\,cm},
 \ee
and $I_{45}=I/10^{45}\,{\rm g\,cm^2}$. If the surface field of the neutron star does not considerably exceed $B_{\rm CRSF}$ the spin-down torque is significantly smaller than $K_0$. In particular, the spin-down torque applied to the neutron star from the Keplerian disk under these conditions ($B_* = B_{\rm CRSF}$) is $K_{\rm sd}^{\rm (d)} = k_{\rm t}\,\mu^2/r_{\rm cor}^3 \sim 2 \times 10^{30}\,k_{\rm t}\,{\rm dyne\,cm}$ (see Eq.~\ref{ksd00}). The spin-up torque in this case can be evaluated as $K_{\rm su} \ga K_0$. Solving this inequality with spin-up torque expressed by Eq.~(\ref{ksugen}) one finds that the spin behavior of the pulsar during the rapid spin-up episodes can be explained in terms of accretion from the Keplerian disk provided the dipole magnetic moment of the neutron star is $\mu \ga \mu_0$, where
 \be\label{mfdisk}
 \mu_0 \simeq 2 \times 10^{30}\,{\rm G\,cm^3}\ I_{45}^{7/2}\ \dmf_{17}^{-3}\ m^{-3/2}\
\left(\frac{\dot{\nu}_{\rm su}}{\dot{\nu}_{\rm u0}}\right)^{7/2}.
 \ee
This corresponds to the surface field $B_0 = 2 \mu_0/R_{\rm ns}^3 \ga 4 \times 10^{12}$\,G, which is close to $B_{\rm CRSF}$.

An attempt to explain the spin-up episodes in terms of the QS accretion scenario encounters difficulties. The spin-up torque in this case could be as high as $K_0$ only if the relative velocity of the neutron star were $v_{\rm rel} \la v_0$, where
 \be
 v_0 \simeq 200\ \xi_{0.2}^{1/4}\ m^{1/2}\ \dmf_{17}^{1/4}\ \left(\frac{P_{\rm orb}}{41.5\,{\rm d}}\right)^{-1/4} {\rm km\,s^{-1}}.
 \ee
However, a formation of the Keplerian disk under these conditions cannot be avoided (see Eq.~\ref{vcr}). It therefore appears that the interpretation of the spin-up episodes in terms of the Keplerian disk accretion scenario is reliable and consistent with the surface field of the neutron star measured through observations of CRSF \citep[for discussion see also][]{Koh-etal-1997}.

   \section{Spin-down trends}\label{spin-down}

The spin-down torque applied to the neutron star during the spin-down trends is $|K_{\rm sd}| = |K_1| + |K_{\rm su}|$, where
 \be\label{k1}
|K_1| = 2 \pi I \dot{\nu}_{\rm sd} \simeq 6 \times 10^{32}\ I_{45} \left(\frac{\dot{\nu}_{\rm sd}}{\dot{\nu}_{\rm d0}}\right) {\rm dyne\,cm}.
 \ee

An interpretation of the pulsar behavior during these trends cannot be made within the existing spin-down scenarios without additional assumptions. If the star accretes material from a Keplerian disk the spin-up torque is $|K_{\rm su}^{\rm (d)}| \gg |K_1|$ (see above). The spin-down behavior of the pulsar in this case implies $|K_{\rm sd}^{\rm (d)}| > |K_{\rm su}^{\rm (d)}|$. To satisfy this condition one has to assume that the surface field of the star is in excess of $5 \times 10^{14}$\,G, which is 100 times larger than $B_{\rm CRSF}$.

The spin-up torque applied to the neutron star is significantly smaller if the star undergoes quasi-spherical accretion. Its value depends on the efficiency of the processes responsible for angular momentum dissipation in the accreting material which is accounted for by the parameter $\xi$ (see Eq.~\ref{ksugen}). This parameter in the case under consideration is limited by the condition $\omega_{\rm en}(r_{\rm m}) < \omega_{\rm s}$, which implies that the angular velocity of the accreting material at the magnetospheric boundary is smaller than the angular velocity of the star itself. Otherwise, the interaction between the accretion flow and stellar magnetosphere tends to spin-up the star \citep[for discussion see,][]{Bisnovatyi-Kogan-1991}. Setting $r_{\rm m} = r_{\rm a}$ and solving the above inequality for the parameters of GX\,301--2 yields
   \begin{eqnarray}
\xi & < & 0.03~m^{-12/7}\ L_{37}^{-4/7}\ R_6^{20/7}\ \left(\frac{P_{\rm orb}}{41.5\,{\rm d}}\right) ~ \times\  \\
    \nonumber
 & & \times\  \left(\frac{P_{\rm s}}{685\,{\rm s}}\right)^{-1} \left(\frac{v_{\rm rel}}{400\,{\rm km\,s^{-1}}}\right)^4 \left(\frac{B_*}{B_{\rm CRSF}}\right)^{8/7}
  \end{eqnarray}
This indicates that $|K_{\rm su}^{(0)}|$ during the spin-down trends does not exceed $|K_1|$ and hence, the condition for the spin-down phase can be expressed as $|K_{\rm sd}| \ga |K_1|$.

Putting $K_{\rm sd} \sim K_{\rm sd}^{(0)}$ one finds that the above condition can be satisfied only if the surface field of the neutron star is $\ga 10^{14}\,k_{\rm t}^{-1/2}$\,G \citep[for discussion see also][]{Lipunov-1982, Doroshenko-etal-2010}. To satisfy the condition $K_{\rm sd}^{\rm (t)} \ga K_1$ one has to assume that either the surface field of the neutron star is $\ga 3\,k_{\rm t}^{-7/8} B_{\rm CRSF}$, or the radius of the stellar magnetosphere is smaller than its canonical value, $r_{\rm a}$, at least by a factor of $2\,k_{\rm t}^{-2/3}$.

Thus, both the Keplerian disk and quasi-spherical accretion scenarios encounters difficulties explaining the spin behavior of GX\,301--2 during the spin-down trends. This may indicate that either the surface field of the star is indeed in excess of $B_{\rm CRSF}$, or the neutron star in this binary system accretes in a different way. The first possibility has been already discussed by \citet{Doroshenko-etal-2010}. Here we focus on the analysis of an alternative accretion scenario. We show that the peculiar spin behavior of GX\,301--2 can be explained in terms of the magnetic accretion scenario \citep{Shvartsman-1971, Bisnovatyi-Kogan-Ruzmaikin-1974, Bisnovatyi-Kogan-Ruzmaikin-1976} provided the material captured by the neutron star from the wind of its normal companion is magnetized.

  \section{Magnetic accretion}\label{mca}

Let us consider a situation in which the magnetic energy density in the material captured by the neutron star, $\msE_{\rm m}(R_{\rm G}) = B_{\rm f0}^2/8 \pi$, is comparable to its thermal energy density, $\msE_{\rm th}(R_{\rm G}) = \rho_{\infty} c_{\rm s0}^2$, i.e. $\beta \equiv \msE_{\rm th}/\msE_{\rm m} \sim 1$. Here $c_{\rm s0}$ is the speed of sound and $B_{\rm f0}$ is the strength of the magnetic field in the captured material at $R_{\rm G}$. The magnetic field in the free-falling material is dominated by its radial component, $B_{\rm r}$, \citep[the transverse scales in the free-falling flow contract as $r^{-2}$, while the radial scales expand as $r^{1/2}$,][]{Zeldovich-Shakura-1969}, which under the magnetic flux conservation condition increases as  $B_{\rm r}(r) \sim B_{\rm f0} \left(R_{\rm G}/r\right)^2$ \citep{Bisnovatyi-Kogan-Fridman1970}. The magnetic pressure in the free-falling material,
\be
 \msE_{\rm m}(r) = \msE_{\rm m}(R_{\rm G}) \left(\frac{R_{\rm G}}{r}\right)^4,
 \ee
increases, therefore, more rapidly than its ram pressure,
 \be
\msE_{\rm ram}(r) = \msE_{\rm ram}(R_{\rm G}) \left(\frac{R_{\rm G}}{r}\right)^{5/2},
 \ee
and hence, the gravitational energy of the star is converted predominantly into the magnetic pressure of the free-falling material, $\msE_{\rm m}/\msE_{\rm ram} \propto r^{-3/2}$. Here $\msE_{\rm ram}(R_{\rm G}) = \rho_{\infty} v_{\rm rel}^2$ is the ram pressure of the captured material at the Bondi radius.

The distance $R_{\rm sh}$ (hereafter Shvartsman radius) at which the magnetic pressure in the accretion flow reaches its ram pressure can be derived by solving equation $\msE_{\rm m}(R_{\rm sh}) = \msE_{\rm ram}(R_{\rm sh})$. This yields \citep{Shvartsman-1971},
 \be\label{rsh}
 R_{\rm sh} = \beta^{-2/3} \left(\frac{c_{\rm s}}{v_{\rm rel}}\right)^{4/3} R_{\rm G} =
 \beta^{-2/3}\ \frac{2 GM_{\rm ns}\,c_{\rm s}^{4/3}}{v_{\rm rel}^{10/3}}.
 \ee

If the magnetic flux in the accreting material is conserved, the accretion ends at the Shvartsman radius. Further accretion in this case is impossible. Otherwise, the magnetic energy in the flow would exceed its gravitational energy, which contradicts the energy conservation law \citep[for discussion see][]{Shvartsman-1971}. Therefore, the accretion flow can approach the star to a closer distance only if dissipation of the magnetic field in the flow occurs. If the field dissipation is governed by magnetic reconnection the characteristic time of the accretion process inside $R_{\rm sh}$ is limited to $t \geq t_{\rm rec}$, where
 \be\label{trec}
 t_{\rm rec} = \frac{r}{\eta_{\rm m} v_{\rm A}} = \eta_{\rm m}^{-1}\ t_{\rm ff}\
 \left(\frac{v_{\rm ff}}{v_{\rm A}}\right).
 \ee
The value of the efficiency parameter $\eta_{\rm m}$ depends on physical conditions and field configuration in the region of reconnection, and in the general case is limited to $0 < \eta_{\rm m} \leq 0.1$ \citep[see, e.g.,][and references therein]{Parker-1971, Kadomtsev-1987, Noglik-etal-2005, Somov-2006}. The Alfv\'en velocity in the accreting material increases as it approaches the neutron star from the initial value $V_{\rm A0} \sim \beta^{-1/2} c_{\rm s}$ to $V_{\rm A}(R_{\rm sh}) \sim V_{\rm ff}$. Hence, the time of the magnetic reconnection in the accretion flow remains significantly larger than the dynamical (free-fall) time, $t_{\rm ff} = r/V_{\rm ff} = \left(r^3/2GM_{\rm ns}\right)^{1/2}$, at any stage of the accretion process.
This proves reliability of the assumption about conservation of the magnetic flux in the free-falling material. But, on the other hand, it also suggests that the accretion flow will be decelerated at the Shvartsman radius by its own magnetic field and the accretion process in the region $r < R_{\rm sh}$ operates in the diffusion approximation.

Rapid amplification of the magnetic field in the spherical flow and deceleration of the flow by its own magnetic field at the Shvartsman radius have been confirmed in analytical studies \citep{Bisnovatyi-Kogan-Ruzmaikin-1974, Bisnovatyi-Kogan-Ruzmaikin-1976} and numerical calculations of magnetized spherical accretion onto a black hole \citep{Igumenshchev-etal-2003, Igumenshchev-2006}. These calculations have shown that the magnetized flow is shock-heated at the Shvartsman radius up to the adiabatic temperature. The structure of the accretion flow inside the Shvartsman radius depends on the efficiency of cooling of the accreting material. If cooling is inefficient the accretion process switches into the convective-dominated stage in which some material is leaving the system in a form of jets \citep{Igumenshchev-etal-2003}. Otherwise, the material tends to flow along the lines of the large scale field of the accreting matter itself and accumulates in a dense non-Keplerian slab \citep[see Fig.\,1 in ][]{Bisnovatyi-Kogan-Ruzmaikin-1976}. The material in the slab is confined by its own magnetic field and its radial motion continues as the field is annihilating. This indicates that the accretion process in the slab occurs on the reconnection timescale expressed by Eq.~(\ref{trec}).

 \subsection{Magnetic accretion in X-ray pulsars}

The Shvartsman radius exceeds the canonical magnetospheric radius of the neutron star evaluated in the quasi-spherical accretion scenario, $r_{\rm a}$, if the relative velocity of the neutron star through the wind of its massive companion satisfies the condition $v_{\rm rel} \leq v_{\rm mca}$, where
 \be
 v_{\rm mca} = \beta^{-1/5}\,(2GM_{\rm ns})^{12/35}\,\mu^{-6/35}\,\dmf^{3/35}\,c_{\rm s}^{2/5}.
 \ee
For typical parameters of HMXBs this velocity is
  \bdm
v_{\rm mca} \simeq 680\ \beta^{-1/5}\ m^{12/35} \mu_{30}^{-6/35}\ \dmf_{17}^{3/35}\ c_6^{2/5}\ {\rm km\,s^{-1}},
 \edm
which under the conditions of interest substantially exceeds the critical velocity at which the formation of the Keplerian disk in the system can be expected, $v_{\rm cr}$ (see Eq.~\ref{vcr}). Here $c_6 = c_{\rm s0}/10^6\,{\rm cm\,s^{-1}}$. This finding allows us to distinguish a subclass of HMXBs in which the accretion occurs in a spherically symmetrical fashion and the accreting material is strongly affected by the magnetic field of the flow itself. This subclass is defined by the condition $v_{\rm cr} < v_{\rm rel} < v_{\rm mca}$.

The structure of the accretion flow inside the Shvartsman radius depends on the efficiency of cooling processes in the accreting material. As shown by \citet{Arons-Lea-1976} and \citet{Elsner-Lamb-1977}, the cooling of the material accreting onto a neutron star is dominated by the inverse Compton scattering of X-ray photos emitted from the stellar surface on the hot electrons of the accretion flow. This mechanism in the considered case will be effective if the Compton cooling time of the accretion flow at the Shvartsman radius, $t_{\rm c}(R_{\rm sh})$, is smaller than the characteristic time of the accretion process, which in the case of the magnetic accretion scenario is the reconnection time, $t_{\rm rec}$. Combining Eqs.~(\ref{tcomp}) and (\ref{trec}), one can express inequality $t_{\rm c}(R_{\rm sh}) \leq t_{\rm rec}$ as $L_{\rm X} \ga L_{\rm cr}$, where
 \be
 L_{\rm cr} \simeq 3 \times 10^{33}\ \mu_{30}^{1/4}\ m^{1/2}\ R_6^{-1/8}\ \left(\frac{\eta_{\rm m}}{0.001}\right) \left(\frac{R_{\rm sh}}{r_{\rm a}}\right)^{1/2}{\rm erg\,s^{-1}}.
 \ee
This indicates that the cooling of magnetized flow can be effective even in faint X-ray pulsars in which the Shvartsman radius does not significantly exceed the Alfv\'en radius of the neutron star. The structure of the accretion flow in those systems in which the above conditions are satisfied can be treated in terms of the magnetic slab.

  \subsection{Magnetospheric radius}

The material in the slab is approaching the neutron star up to a distance at which its gas pressure becomes equal to the magnetic pressure due to the stellar dipole magnetic field. The gas density in the slab at this distance (the magnetospheric radius) is, therefore,
  \be\label{rhosl}
  \rho_{\rm sl} = \frac{\mu^2\,m_{\rm p}}{2 \pi\,k_{\rm B}\,T_0\,r_{\rm m}^6},
  \ee
where $T_0$ is the temperature of the material at the inner radius of the slab.

The value of the magnetospheric radius in the general case depends on the mode by which the material enters the stellar field at the magnetospheric boundary. As shown by \citet{Elsner-Lamb-1984}, the diffusion rate of the accreting material into the stellar field can be evaluated as
    \begin{eqnarray}\label{dmfin-1}
 \dmf_{\rm in}(r_{\rm m}) & = & 4 \pi r_{\rm m} \delta_{\rm m} \rho_0 V_{\rm ff}(r_{\rm m}) \\
   \nonumber
  & = & 4 \pi r_{\rm m}^{5/4} D_{\rm eff}^{1/2} \rho_0 (2 GM_{\rm ns})^{1/4},
   \end{eqnarray}
where $\delta_{\rm m} = \left(D_{\rm eff}\ \tau_{\rm d}\right)^{1/2}$ is the thickness of the diffusion layer at the magnetospheric boundary (magnetopause) and $D_{\rm eff}$ is the effective diffusion coefficient. The diffusion time is determined by the time on which the material being penetrated into the field leaves the magnetopause by free-falling along the magnetospheric field lines towards the stellar surface, $\tau_{\rm d} \sim t_{\rm ff}(r_{\rm m})$.

The estimate~(\ref{dmfin-1}) has good theoretical and observational grounds. It has been justified by studies of the Earth's magnetosphere which show that the rate of solar wind penetration into the magnetic field of the Earth can be evaluated taking $D_{\rm eff} \sim D_{\rm B}$, where
 \be\label{dbohm}
D_{\rm B} = \alpha_{\rm B} \frac{c k_{\rm B} T_0 r_{\rm m}^3}{2 e \mu}
 \ee
is the Bohm diffusion coefficient, $e$ is the electron charge and $\alpha_{\rm B}$ is the efficiency parameter, which ranges in $0.1-0.25$ \citep{Gosling-etal-1991}. The diffusion process in this case is governed by magnetic reconnection and drift-dissipative instabilities \citep[][and references therein]{Paschmann-2008}. Note also that the same diffusion rate has previously been measured in plasma experiments with TOKAMAKs \citep{Kadomtsev-Shafranov-1983} and evaluated from observations of solar flares \citep{Priest-1982}.

A higher rate of plasma penetration into the magnetosphere could be expected if the magnetospheric boundary were interchange unstable. The Rayleigh-Tailor instability of the boundary can occur in bright long-period X-ray pulsars. This instability would be, however, suppressed by the magnetic pressure gradient if the X-ray luminosity of the neutron star undergoing spherical accretion is $L_{\rm X} < 3 \times 10^{36}\,{\rm erg\,s^{-1}}$ \citep{Arons-Lea-1976, Elsner-Lamb-1977}, and by the magnetic field shear in the magnetopause if the spin period of the neutron star is smaller than a few hundred seconds \citep[see e.g.][and references therein]{Ikhsanov-Pustilnik-1996}. The Kelvin-Helmholtz instability can be effective in short-period X-ray pulsars in which the magnetospheric radius of the neutron star is close to its corotation radius \citep{Burnard-etal-1983}. But this instability might not be effective if the relative velocity between the magnetosphere and the accretion flow exceeds the speed of sound or/and if the material at the boundary is magnetized \citep[for discussion see][]{Anzer-Boerner-1980, Anzer-Boerner-1983, Malagoli-etal-1996}. This indicates that the interchange instabilities of the boundary can be at work only in a limited number of pulsars, while Bohm diffusion takes place in any of the considered objects. Having this in mind, we focus our consideration on the case of Bohm diffusion assuming that the interchange instabilities of the boundary are suppressed.

The stationary accretion picture implies that the rate of plasma diffusion into the pulsar field at the magnetospheric boundary is equal to the mass capture rate by the neutron star from its environment at the Bondi radius and to the mass accretion rate onto the surface of the neutron star. Let us assume that this condition is satisfied at the distance $r_{\rm mca}$, i.e. $\dmf_{\rm in}(r_{\rm mca}) = L_{\rm X} R_{\rm ns}/GM_{\rm ns}$. Combining Eqs.~(\ref{rhosl}), (\ref{dmfin-1}) and (\ref{dbohm}), and solving the above equation for $r_{\rm mca}$, one gets
 \be\label{rmb}
r_{\rm mca} \simeq 8 \times 10^7\ \alpha_{0.1}^{2/13} \mu_{30}^{6/13} T_6^{-2/13} m^{5/13} R_6^{-4/13}  L_{37}^{-4/13}\,{\rm cm}
  \ee
Here $\alpha_{0.1}=\alpha/0.1$, $T_6 = T_0/10^6$\,K is the plasma temperature at the magnetospheric boundary, which is normalized according to \citet{Masetti-etal-2006}, and $L_{37}$ is the X-ray luminosity of the pulsar in units of $10^{37}\,{\rm erg\,s^{-1}}$.

The value of $r_{\rm mca}$ under the same conditions is smaller than the value of $r_{\rm a}$, which represents the canonical radius of the magnetosphere of the neutron star undergoing spherical accretion. This can be associated with accumulation of material at the inner radius of the slab which occurs if $\dmf_{\rm in} < \dmf_{\rm c}$. As the material accumulates the gas pressure increases and the slab approaches the neutron star to a closer distance. Since $\dmf_{\rm in} \propto r_{\rm m}^{-13/4}$ the diffusion rate of material into the magnetosphere also increases and reaches $\dmf_{\rm c}$ as the inner radius of the slab decreases to $r_{\rm mca}$.

The spin-down torque applied to the neutron star from the magnetic slab can be evaluated as $K_{\rm sd}^{\rm (sl)} = k_{\rm t} \dmf \omega_{\rm s} r_{\rm mca}^2$. Taking into account that $\dmf = \mu^2/ \left(2 GM_{\rm ns} r_{\rm mca}^7 \right)^{1/2}$ one finds
 \be
K_{\rm sd}^{\rm (sl)} = \frac{k_{\rm t}\,\mu^2\,\omega_{\rm s}}{r_{\rm mca}^{3/2} (2 GM_{\rm ns})^{1/2}} =  \frac{k_{\rm t}\,\mu^2}{\left(r_{\rm mca}\,r_{\rm cor}\right)^{3/2}}
 \ee
Thus, the spin-down torque applied to the neutron star accreting material from the magnetic slab is a factor of $\sim \left(r_{\rm a}/r_{\rm mca}\right)^{3/2}$ higher than the maximum possible spin-down torque applied to the star accreting material at the same rate from a quasi-spherical flow.

 \section{Signs of magnetic accretion in GX\,301--2}\label{magx}

The maximum possible spin-down rate of the neutron star in GX\,301--2 within the magnetically controlled accretion scenario can be evaluated as $\dot{\nu}_{\rm sd}^{\rm (mca)} = K_{\rm sd}^{\rm (sl)}/2 \pi I$, that for the pulsar parameters is
  \begin{eqnarray}
\dot{\nu}_{\rm sd}^{\rm (mca)} & \simeq & 7 \times 10^{-13}\,{\rm Hz\,s^{-1}} ~ k_{\rm t}\ \alpha_{0.1}^{-3/13}\ m^{-14/13} \\
     \nonumber
 &   \times\ & I_{45}^{-1}\ T_6^{3/13}\ L_{37}^{6/13} R_6^{57/13}\ P_{685}^{-1}\ \left(\frac{B_*}{B_{\rm CRSF}}\right)^{17/13}
  \end{eqnarray}
Hence, the spin-down rate of the neutron star observed during the spin-down trends can be explained in terms of the magnetically controlled accretion scenario provided the surface field of the star is close to $B_{\rm CRSF}$ and the efficiency coefficient is $k_{\rm t} \geq 0.14$. The spin-down power of the neutron star within this scenario is spent in the energy of electric currents and turbulent motions excited at the magnetospheric boundary and, possibly, to spin-up the material at the inner radius of the magnetic slab. In this case the angular momentum lost by the star can partly be accumulated in the material surrounding its magnetosphere. The spin-down torque in this case decreases as the angular velocity of the material at the inner radius of the slab approaches the angular velocity of the star itself. The angular momentum accumulated in the slab can later be transferred back to the star due to accretion process. The star in this case is expected to switch from the spin-up to the spin-down phase without any significant change of the accretion flow geometry.

A detailed study of the angular momentum exchange between the star and the slab is beyond the scope of this paper. Here we would like to note only that the angular velocity to which the material in the slab could be spun-up within this scenario is limited to the angular velocity of the star itself, $\omega_{\rm s}$. The angular momentum transfer from the star to the slab can bring its inner part to a solid rotation. The radial size of this part is limited, however, to the corotation radius at which $\omega_{\rm s} = \omega_{\rm k}(r_{\rm cor})$. This opens a possibility to consider a situation in which the spin-up of the material at the inner radius of the slab results in temporary transition of the accretion picture from the non-Keplerian slab to Keplerian disk of the radius $r_{\rm cor}$. The star in this case will experience a rapid spin-up on a timescale of the viscous time at the corotation radius. It is interesting that the size of the Keplerian disk evaluated by \citet{Koh-etal-1997} from modelling the rapid spin-up events is very close to the corotation radius of the pulsar.

  \section{Discussion}\label{discussion}

One of the basic assumptions adopted in the magnetically controlled accretion scenario is that the magnetic field energy density in the material captured by the neutron star at the Bondi radius is comparable to its thermal energy density. The latter can be evaluated by taking into account that $\rho_{\infty} = \dmf/\pi R_{\rm G}^2 v_{\rm rel}$. This yields
 \be
\msE_{\rm th} \simeq 0.02\ {\rm erg\,cm^{-3}}\ m^{-2}\ c_6^2\ \dmf_{17}\ \left(\frac{v_{\rm rel}}{500\,{\rm km\,s^{-1}}}\right)^3.
 \ee
The magnetic energy density in the captured wind depends on the field strength of the massive companion of the neutron star. The dipole approximation ($B_{\rm mc}  \propto a^{-3}$) to the magnetic field of the massive stars remains valid up to a distance $a_{\rm k}$, at which the dynamical pressure of the wind ejecting by this star reaches the magnetic tension of the stellar dipole field (here $a$ is the distance from the massive companion). The magnetic field in the material propagating beyond $a_{\rm k}$ decreases as $B \propto a^{-2}$ \citep{Walder-etal-2011}. Therefore, the magnetic energy density in the wind at the binary separation $a_0 > a_{\rm k}$ is $\msE_{\rm m}(a_0) = \msE_{\rm m0} = \mu_{\rm ms}^2/\left(2 \pi a_{\rm k}^2 a_0^4\right)$, which implies
  \be\label{bwa}
\msE_{\rm m0} \simeq 0.33\,{\rm erg\,cm^{-3}}\,a_{13}^{-4}
 \left(\frac{\mu_{\rm ms}}{10^{39}\,{\rm G\,cm^3}}\right)^2 \left(\frac{a_{\rm k}}{100\,R_{\sun}}\right)^{-2}
 \ee
Here $\mu_{\rm ms}$ is the dipole magnetic moment of the massive star and $a_{13}$ is the binary separation in units of $10^{13}$\,cm. It, therefore, appears, that the magnetic accretion scenario can be expected in GX\,301--2 if the surface field of the massive component is a few hundred Gauss.

Recent spectropolarimetric observations \citep[see, e.g.][and references therein]{Hubrig-etal-2006, Oksala-etal-2010, Martins-etal-2010} have shown a relatively strong magnetization of O/B-type stars to be not unusual. The strength of the large-scale field at the surface of these objects has been measured in the range $\sim 500-5000$\,G, and in some cases beyond 10\,kG. In this light, the assumption about relatively strong magnetization of the massive star in GX\,301--2 seems to be rather reliable. It can be tested by spectropolarimetric observations of this system, which, therefore, appears to be of great importance for justification of the magnetically controlled accretion scenario used in our paper.

  \section{Conclusion}\label{concl}

We have shown that the rapid spin-down of the neutron star observed during the spin-down trends of GX\,301--2 can be explained in terms of magnetically controlled accretion scenario provided the surface field of the star is close to the value, $B_{\rm CRSF}$, derived from observations of the cyclotron resonance scattering feature in the X-ray spectrum of this pulsar. The same strength of the magnetic field has been evaluated from modeling of the rapid spin-up events in terms of the Keplerian disk accretion. We conclude that the problems in modeling of the spin evolution of GX\,301--2 encountered by \citet{Doroshenko-etal-2010}  indicate an oversimplification in the currently used accretion scenarios rather then an extremely high magnetization of the neutron star. We have shown that these problems can be avoided by incorporation of the magnetic filed of the accretion flow into the quasi-spherical accretion scenario provided the relative velocity of the neutron star satisfies the condition $v_{\rm cr} < v_{\rm rel} < v_{\rm mca}$. The accretion flow in this case is decelerated by its own magnetic field at the Shvartsman radius, which exceeds the magnetospheric radius of the neutron star, and accumulates in the non-Keplerian magnetic slab confined by the magnetic field of the flow itself. The plasma approaches the star on the timescale of field dissipation in the magnetic slab, which significantly exceeds the dynamical (free-fall) timescale. The process of plasma penetration into the stellar magnetosphere is governed by the magnetic reconnection and drift-dissipative instabilities and occurs at a rate of the Bohm diffusion. The magnetospheric radius within this scenario, $r_{\rm mca}$, is smaller than the canonical value, $r_{\rm a}$. The spin-down torque applied to the neutron star from the slab is $K_{\rm sd} = k_{\rm t}\,\mu^2/\left(r_{\rm mca} r_{\rm cor}\right)^{3/2}$, where $0.1 < k_{\rm t} <1$. An exchange of the angular momentum between the star and the slab can lead to variations of the spin period of the pulsar.

\acknowledgments

We would like to thank G.S.\,Bisnovatyi-Kogan, L.A.\,Pustil'nik  and N.G.\,Beskrovnaya for useful discussions and suggesting improvements. M.H.F. acknowledges support from NASA grant NNX11AE24G. The research has been also supported by the Program of Presidium of Russian Academy of Sciences N\,21, and NSH-1625.2012.2.

\end{document}